\documentclass[12pt,english,floatfix,nofootinbib,superscriptaddress,aps,prd,preprint]{revtex4}
\usepackage[utf8]{inputenc}
\usepackage{float}
\usepackage{array}
\usepackage{lipsum}
\usepackage{graphicx}
\usepackage{amsmath,amsthm,amsfonts,amssymb}
\usepackage{graphicx}
\usepackage[english]{babel}
\usepackage{color}
\usepackage{subfig}
\usepackage{caption}
\usepackage{tensor}
\usepackage{esint}
\usepackage[dvips]{epsfig}
\usepackage[dvips]{graphicx}
\usepackage{float}
\usepackage{units}
\usepackage{textcomp}
\usepackage{mathrsfs}
\usepackage{amsmath}
\usepackage[makeroom]{cancel}
\usepackage{amssymb}
\usepackage{amsbsy}
\usepackage{amsfonts}
\usepackage{amssymb,mathrsfs,xcolor}
\usepackage{esint}
\usepackage{array}
\usepackage{graphicx}

\usepackage{hyperref}
\hypersetup{
    colorlinks,
    citecolor=blue,
    filecolor=green,
    linkcolor=purple,
    urlcolor=red,
}

\usepackage{slashed}

\newcommand{\ie}{\begin{equation}}
\newcommand{\fe}{\end{equation}}
\newcommand{\se}{\begin{eqnarray}}
\newcommand{\ff}{\end{eqnarray}}

\usepackage{hyperref}
\hypersetup{colorlinks,breaklinks,
			citecolor=[rgb]{0,0.0,1.0},
            urlcolor=[rgb]{0.0,0.0,1.0},
            linkcolor=[rgb]{0,0.5,0.9}}

\begin{document}

\title{G\"{o}del-type solutions within the $f(R,Q,P)$ gravity}

\author{J. R. Nascimento, A. Yu. Petrov, P. J. Porf\'{i}rio, Ramires N. da Silva}
\affiliation{Departamento de F\'{\i}sica, Universidade Federal da Para\'{\i}ba\\
 Caixa Postal 5008, 58051-970, Jo\~ao Pessoa, Para\'{\i}ba, Brazil}
\email{jroberto, petrov, pporfirio@fisica.ufpb.br, rns2@academico.ufpb.br}




\date{\today}

\begin{abstract}
In this work, we scrutinize the consistency of spacetime homogeneous G\"{o}del-type metrics within $f(R,Q,P)$ theories of gravity for well-motivated matter sources. As it is well known, such geometries allow for causality violation. We provide general conditions to engender completely causal solutions in a manner completely different from general relativity. We take some specific models, for instance, $f(R,Q,P)=R -\dfrac{\mu^{4n+2}}{\left(a R^2+b Q+c P\right)^n}$, to illustrate the general results. Notably, we also find an unusual completely causal vacuum solution in the presence of a non-trivial cosmological constant  which corresponds to the case $m^2=4\omega^2$.

\end{abstract}

\maketitle

\section{Introduction}

Studies concerning the nature of gravitational interaction have become one of the most prominent topics in modern theoretical physics, opening the door to whether general relativity (GR) is the correct theory to describe gravity at different energy scales. At the same time, experimental observations show two different gravitational phenomena: the late-time accelerated expansion of the Universe and the discrepancy between the experimental and theoretical data of the rotation curves of galaxies in clusters \cite{Riess}, which cannot be explained in the GR framework without the deploy of dark matter and energy \cite{CarCos}.
Another motivation being essentially important at the quantum level, that is, the non-renormalizability of Einstein's theory, called attention to modified gravity models. Further, the increase of interest to gravity was supported by the detection of gravitational waves \cite{LIGO1, LIGO2, LIGO3}, obtaining the first image of the black hole \cite {Akiyama}, and further, obtaining the image of the shadow of a black hole in the center of the Milky Way \cite{AkiyamaBH}.

Historically, the first interesting classical results within modified gravity have been obtained already in \cite{Staro} where the possibility of de Sitter cosmological solutions displaying exponential growth was proved for the modified Einstein equations involving terms of the second order in curvature. Further, various extensions of Einstein's gravity were proposed. The class of possible modifications of the gravity Lagrangian is very wide -- they include the use of a generic function of the scalar curvature (see f.e. \cite{Sotiriou} and references therein) or other manners to implement higher derivatives in a pure gravitational sector, adding of extra fields (scalar, vector or even tensor ones) which, instead of being treated as a matter, are considered as ingredients of a complete description of gravity (for a general review on modified gravity, see \cite{Nojiri:2010wj,Nojiri:2017ncd,ourrev}). 

Among various manners to modify purely gravitational sector, very interesting ones are based on the use of higher curvature invariants, namely, $Q=R_{\mu\nu}R^{\mu\nu}$ and $P=R_{\mu\nu\rho\sigma}R^{\mu\nu\rho\sigma}$, so, one can introduce $f(R,Q)$ or $f(R,Q,P)$ gravities, which give origin to a more wide class of extended gravity theories, where, certainly, new solutions impossible in general relativity can arise. Detailed studies of various issues related to $f(R,Q)$ gravities, both in metric and Palatini formulations, have been performed earlier, see f.e. \cite{frq} and references therein. So, the natural extension of this study consists in treating $f(R,Q,P)$ theories. Originally, this class of theories was proposed in \cite{CappLau2009} where some of its applications within the gravitational wave context were discussed, further, a detailed discussion of these theories (especially, their specific form of $f(R,\mathcal{G})$ gravities, with $\mathcal{G}$ is the Gauss-Bonnet invariant) within the cosmological context was presented in \cite{OOB, Elizalde:2010jx, Bamba:2010wfw, DeLaurentis:2015fea, Benetti:2018zhv, DeFelice:2010sh, delaCruz-Dombriz:2011oii}.  Therefore, the study of the consistency of other known solutions of GR within these theories is very natural. 

Within this paper, we concentrate on the class of G\"{o}del-type metrics known to generate closed timelike curves (CTCs) for certain values of their parameters \cite{RT, RT2, GOD}. Earlier, the consistency of these metrics, besides GR, was checked, together with verifying possibilities of causal solutions, also within Chern-Simons modified gravity \cite{CS1,CS2}, Brans-Dicke gravity \cite{BD}, $f(R,Q)$ gravity \cite{frq} and some other gravity models. The aim of this paper consists of checking the validity of G\"{o}del-type solutions in $f(R,Q,P)$ gravity. We generalize, in some sense, the results found in \cite{barr} 
 since we are considering all three classes of G\"{o}del-type metrics while only the hyperbolic class is taken into account in \cite{barr}. Furthermore, we derive expressions for the critical radius for the linear and hyperbolic classes of G\"{o}del-type metrics.  

The structure of the paper looks like follows. In Section \ref{secG}, we describe our model and write down the field equations. In Section \ref{secgo}, we present a brief review on the main properties of G\"{o}del-type metrics. In Section \ref{secsol}, we find G\"{o}del-type solutions for well-motivated matter sources within the framework of $f(R,Q,P)$ theories of gravity. Finally, we give some conclusions and a summary.

\section{The setup: $f(R,Q,P)$ gravity}\label{secG}
\subsection{Action and field equations}
We start this section by introducing the $f(R,Q,P)$ gravity action (see f.e. \cite{CappLau2009}):
\begin{equation}
    S=\frac{1}{2\kappa^2}\int d^{4}x\,\sqrt{-g}f(R,Q,P)+\int d^{4}x\,\sqrt{-g}\mathcal{L}_{mat}(g_{\mu\nu},\psi),
    \label{action}
\end{equation}
where $\kappa^2=8\pi G$ is related to Newton's constant $G$, $g$ is the determinant of the spacetime metric $g_{\mu\nu}$, $f(R,Q,P)$ is a generic function of the following scalar geometrical invariants: the Ricci scalar, $R= g^{\mu\nu}R_{\mu\nu}$, $Q=R^{\mu\nu}R_{\mu\nu}$ and $P=R^{\alpha}_{\,\,\,\beta\mu\nu}R_{\alpha}^{\,\,\,\beta\mu\nu}$. Furthermore, $\mathcal{L}_{mat}$ is the Lagrangian of the matter sources represented above by the arbitrary fields $\psi$. 

The gravitational field equations are obtained by the variation of the action (\ref{action}) with respect to the metric. Then, by doing so, one gets 
\begin{eqnarray}
& & \nonumber f_R R_{\mu\nu}- \frac{f}{2} g_{\mu\nu} +2 f_Q R^{\beta}_{(\mu} R_{\nu)\beta} + g_{\mu\nu} \Box f_R - \nabla_{(\mu} \nabla_{\nu)} f_R + \Box (f_Q R_{\mu\nu}) - 2 \nabla_{\lambda} \left[\nabla_{(\mu} (f_Q R^\lambda_{\nu)})\right]+\\
&+& g_{\mu\nu} \nabla_{\alpha} \nabla_{\sigma} (f_Q R^{\alpha\sigma}) + 2f_P R_{\alpha\beta\sigma\mu} R^{\alpha\beta\sigma}_{\,\,\,\,\,\,\,\,\,\,\nu} - 4 \nabla_\alpha \nabla_\beta[f_P R^{\alpha\,\,\,\,\,\,\,\,\,\beta}_{\,\,\,(\mu\nu)}] = \kappa^2 T_{\mu\nu}^{(m)},
\label{fe}
\end{eqnarray}
   where the covariant d'Alembertian operator is as usual defined as $\Box\equiv g^{\mu\nu}\nabla_{\mu}\nabla_{\nu}$  and the other quantities are defined as follows: $f_{R}\equiv \frac{\partial f}{\partial R}$, $f_{Q}\equiv \frac{\partial f}{\partial Q}$  and $f_{p}\equiv \frac{\partial f}{\partial P}$. On the r.h.s. of Eq. (\ref{fe}), the stress-energy tensor coming from the contributions of the matter sources is explicitly given by $T_{\mu\nu}^{(m)}=-\frac{2}{\sqrt{-g}}\frac{\delta\left(\sqrt{-g}\mathcal{L}_{mat}\right)}{\delta g^{\mu\nu}}$.

   The above field equations can be conveniently rewritten in a compact shape, namely,
   \begin{equation}
G_{\mu\nu}=\kappa^2_{\text{eff}}T_{\mu\nu}^{(m)}+T_{\mu\nu}^{\text{eff}},
\label{eqw}
   \end{equation}
with $\kappa_{\text{eff}}^2\equiv \frac{\kappa^2}{f_R}$ and
\begin{equation}
\begin{split}
    T_{\mu\nu}^{\text{eff}}&=\frac{1}{f_{R}}\bigg(-\frac{1}{2}Rg_{\mu\nu}f_{R}+\frac{f}{2}g_{\mu\nu}-2 f_Q R^{\beta}_{(\mu} R_{\nu)\beta}-g_{\mu\nu} \Box f_R+\nabla_{(\mu} \nabla_{\nu)} f_R - \Box (f_Q R_{\mu\nu}) +\\
    &+2 \nabla_{\lambda} \left[\nabla_{(\mu} (f_Q R^\lambda_{\nu)})\right]-g_{\mu\nu} \nabla_{\alpha} \nabla_{\sigma} (f_Q R^{\alpha\sigma})-2f_P R_{\alpha\beta\sigma\mu} R^{\alpha\beta\sigma}_{\,\,\,\,\,\,\,\,\,\,\nu} +\\
    &+ 4 \nabla_{\alpha} \nabla_{\beta}[f_P R^{\alpha\,\,\,\,\,\,\,\,\,\beta}_{\,\,\,(\mu\nu)}]\bigg),
    \end{split}
\end{equation}
where $T_{\mu\nu}^{\text{eff}}$ is the effective stress-energy tensor.

Now, taking the trace of Eq.(\ref{eqw}), one finds
\begin{equation}
    R=-\left(\kappa^2_{\text{eff}}T^{(m)}+T^{\text{eff}}\right),
    \label{cons}
\end{equation}
where  the shorthand notation $T^{(m)}=g^{\mu\nu}T_{\mu\nu}^{(m)}$ and $T^{\text{eff}}=g^{\mu\nu}T_{\mu\nu}^{\text{eff}}$ has been used. Plugging this back into Eq.(\ref{eqw}), one arrives at the trace-reversed form of the field equations,
\begin{equation}
R_{\mu\nu}=\kappa^2_{\text{eff}}\left(T_{\mu\nu}^{(m)}-\frac{1}{2}g_{\mu\nu}T^{(m)}\right)+\left(T_{\mu\nu}^{\text{eff}}-\frac{1}{2}g_{\mu\nu}T^{\text{eff}}\right).
\label{eqfi}
\end{equation}
It is straightforward to note from the previous gravitational field equations that they, in general,  contain  higher-order derivative terms, more precisely, fourth-order ones. In this respect, it is well known that these models could present unpleasant pathological behavior, such as the arising of ghost-like instabilities, then leading to the breakdown of the unitarity, see f.e. \cite{HawkingUn}. To avoid this problem, in this paper we shall follow the methodology of the effective field theories \cite{Georgi} which means that the effects of the high curvature terms are suppressed by a typical high energy scale; thus, to some extent, such instabilities can be neglected at the low energy limit.

\section{G\"{o}del-type Metrics }\label{secgo}

Our aim in this section is to provide a brief review of the main features of G\"{o}del-type metrics, in particular, those fulfilling the conditions of space-time homogeneity (ST-homogeneous). We also discuss the causality aspects of such metrics and their three distinct classes. 

The explicit form of the line element of the G\"{o}del-type metrics in cylindrical coordinates looks like \cite{RT, RT2}
\begin{equation}
    ds^{2}=-[dt+H(r)d\theta]^{2}+D^{2}(r)d\theta^{2}+dr^{2}+dz^{2}\label{goty_ds2},
\end{equation}
where $H(r)$ and $D(r)$ are arbitrary functions of the radial coordinate. However, it was proved in \cite{RT} that 
 to attain ST-homogeneity of G\"{o}del-type metrics, the metric functions must satisfy the following conditions (necessary and sufficient)
\begin{equation}
\begin{array}{ccc}
\dfrac{H'(r)}{D(r)} & = & 2\omega,\\
\dfrac{D''(r)}{D(r)} & = & m^{2},
\end{array}\label{eq:st_hom}
\end{equation}
where the prime stands for derivative with respect to the radial coordinate, $r$. The pair $(m^{2},\omega)$ completely characterizes all ST-homogeneous G\"{o}del-type metrics. Physically, the parameter $\omega$ is the vorticity, while the other parameter is allowed to assume any real value, $-\infty\leq m^{2}\leq\infty$. Henceforth, we shall restrict our analysis  to ST-homogeneous G\"{o}del-type metrics only and refer to them simply for G\"{o}del-type metrics. The solutions of Eq.(\ref{eq:st_hom}) define three different classes of  G\"{o}del-type metrics for $\omega\neq 0$, depending on the sign of the parameter $m^2$, namely,

i)\textit{ hyperbolic class}, where $m^{2}>0$ and
\begin{equation}
\begin{array}{cl}
H(r) & =\frac{2\omega}{m^{2}}[\cosh(mr)-1],\\
D(r) & =\frac{1}{m}\sinh(mr);
\end{array}\label{eq:goty_hip}
\end{equation}

ii)\textit{ trigonometric class}, where $m^{2}=-\mu^{2}<0$ and
\begin{equation}
\begin{array}{cl}
H(r) & =\frac{2\omega}{\mu^{2}}[1-\cos(\mu r)],\\
D(r) & =\frac{1}{\mu}\sin(\mu r);
\end{array}\label{eq:goty_trig}
\end{equation}

iii)\textit{ linear class}, where $m^{2}=0$ and
\begin{eqnarray}
H(r) & = & \omega r^2,\label{eq:goty_lin}\\
D(r) & = & r.\nonumber 
\end{eqnarray}
The case corresponding to $\omega=0$ is the degenerate class and it will not take into account here. Note that the well-known G\"{o}del metric \cite{GOD}, which is a solution of Einstein equations with a cosmological constant  $\Lambda$ supported by a dust of density $\rho$, is achieved by taking $m^2=2\omega^2=-2\Lambda=\kappa^2 \rho$. Therefore, the G\"{o}del metric is an example of the hyperbolic class. Regarding isometry groups, the G\"{o}del-type metrics admit different groups of isometries by depending on the relations between $m^2$ and $\omega^2$. As an example, the special class $m^2=4\omega^2$ admits  $G_7$ as the isometry group, which is the larger one.

One of the most interesting properties of the G\"{o}del-type metrics is the presence of closed time-like curves (CTCs) which are defined by circles $C=\lbrace(t,r,\theta,z);\,t,r,z=const,\,\theta\in[0,2\pi]\rbrace$,
in a region restricted by the range $r_{1}<r<r_{2}$, where the function $G(r)=D^{2}(r)-H^{2}(r)$ must be negative inside this region. For the linear class $m=0$, there exists a non-causal region $r>r_{c}$
possessing  closed time-like curves, where 
\begin{equation}
r_{c}=1/\omega
\label{erf}
\end{equation}
is the critical
radius. For the trigonometric class $m^{2}=-\mu^{2}<0$, there exists
an infinite sequence of alternating causal and non-causal regions.
For the hyperbolic class of such spacetimes, there exists a non-causal region $r>r_{c}$, where the critical radius $r_{c}$
is given by
\begin{equation}
\sinh^{2}\left(\dfrac{mr_{c}}{2}\right)=\left(\dfrac{4\omega^{2}}{m^{2}}-1\right)^{-1},\label{eq:raiocr}
\end{equation}
when $0<m^{2}<4\omega^{2}$. Note, however, that from the above equation, when $m^{2}\geq4\omega^{2}$, the presence of CTCs is totally avoidable, that is, there is no breakdown of causality. In the special case $m^{2}=4\omega^{2}$, the critical radius $r_{c}\rightarrow\infty$
\cite{RT}. 

\section{G\"{o}del-type solution in $f(R,Q,P)$ gravity}\label{secsol}

Having found the gravitational field equations in Section \ref{secG}, we now check the viability of G\"{o}del-type metrics within $f(R,Q,P)$ gravity. To begin with, it is convenient to adopt a local Lorentz co-frame $\theta^{A}=e_{\;\;\mu}^{A}dx^{\mu}$ \footnote{Here, capital Latin indices label coordinate fibers of the vector bundle with  structure group $SO(3,1)$, while small Greek indices label, as usual, spacetime coordinates.} for the metric (\ref{goty_ds2}). In particular, we pick the following choice 
\begin{equation}
\begin{split}\theta^{(0)} & =dt+H(r)d\theta,\\
\theta^{(1)} & =dr,\\
\theta^{(2)} & =D(r)d\theta,\\
\theta^{(3)} & =dz.
\end{split}
\label{eq:Lor_tetr}
\end{equation}
Then the line element can be cast into the form:
\begin{equation}
    ds^2 =\eta_{AB}\theta^{A}\theta^{B},
\end{equation}
where $\eta_{AB}=\text{diag}(+1,-1,-1,-1)$ is the Minkowski metric. In the local Lorentz (co)-frame (\ref{eq:Lor_tetr}), the field equations (\ref{eqfi}) read
\begin{equation}
R_{AB}=\kappa^2_{\text{eff}}\left(T_{AB}^{(m)}-\frac{1}{2}\eta_{AB}T^{(m)}\right)+\left(T_{AB}^{\text{eff}}-\frac{1}{2}\eta_{AB}T^{\text{eff}}\right),
\end{equation}
where local Lorentz indices are converted into spacetime ones by means of the vierbein $e^{A}_{\;\;\mu}$ or its inverse $e^{\;\;\mu}_{A}$. For example, the rule to get the Ricci tensor in the co-frame basis (\ref{eq:Lor_tetr}) is $R_{AB}=e^{\;\;\mu}_{A}e^{\;\;\nu}_{B}R_{\mu\nu}$. The specific choice (\ref{eq:Lor_tetr}) is useful since the non-vanishing components of the Ricci tensor, in this frame, are,
\begin{eqnarray}
    R_{(0)(0)}=2\omega^2,\,\, R_{(1)(1)}=R_{(2)(2)}=2\omega^2 - m^2,
\end{eqnarray}
which, in turn, are constant quantities. On top of that, the scalar quantities are also constant as we can check by  direct computation. So, 
\begin{eqnarray}
    R&=&2(m^2-\omega^2);\\
    Q&=&2m^2 (m^2 - 4\omega^2)+12\omega^4;\\
    P&=& 4m^2 (m^2 -6\omega^2) + 44\omega^4.
\end{eqnarray}
Taking this into consideration, the effective stress-energy  tensor reduces to
\begin{equation}
\begin{split}
    T_{\mu\nu}^{\text{eff}}&=\frac{1}{f_{R}}\bigg(-\frac{1}{2}Rg_{\mu\nu}f_{R}+\frac{f}{2}g_{\mu\nu}-2 f_Q R^{\beta}_{(\mu} R_{\nu)\beta} - f_{Q}\Box R_{\mu\nu} +
    2 f_Q \nabla_{\lambda} \left[\nabla_{(\mu}  R^\lambda_{\nu)}\right]-\\
    &-f_{Q}g_{\mu\nu} \nabla_{\alpha} \nabla_{\sigma} R^{\alpha\sigma}-2f_P R_{\alpha\beta\sigma\mu} R^{\alpha\beta\sigma}_{\,\,\,\,\,\,\,\,\,\,\nu} + 4 f_{P}\nabla_{\alpha} \nabla_{\beta} R^{\alpha\,\,\,\,\,\,\,\,\,\beta}_{\,\,\,(\mu\nu)}\bigg),
    \end{split}
\end{equation}
 which can be simplified further by using the following identities
 \begin{equation}
     \begin{split}
         \nabla_{\mu}\nabla_{\nu}R^{\mu\nu}&=\frac{1}{2}\Box R;\\
\nabla_{\rho}\nabla_{\nu}R^{\rho}_{\,\,\mu}&=\frac{1}{2}\nabla_{\nu}\nabla_{\mu}R+R_{\mu\lambda\theta\nu}R^{\lambda\theta}+R^{\lambda}_{\,\,\mu}R_{\nu\lambda};\\
\nabla_{\beta}\nabla_{\alpha}R_{\mu\;\;\;\;\nu}^{\;\;\alpha\beta}&=-\Box R_{\mu\nu}+\nabla_{\beta}\nabla_{\nu}R_{\mu}^{\;\;\beta},
     \end{split}
 \end{equation}
then
\begin{equation}
\begin{split}
     T_{\mu\nu}^{\text{eff}}&=\frac{1}{f_{R}}\bigg(-\frac{1}{2}Rg_{\mu\nu}f_{R}+\frac{f}{2}g_{\mu\nu} - \left(f_{Q}+4f_{P}\right)\Box R_{\mu\nu} +
    2 \left(f_Q +2f_{P}\right) R_{(\nu\;\;\;\mu)}^{\;\;\;\lambda\theta}R_{\lambda\theta}-\\
    &-2f_P R_{\alpha\beta\sigma(\mu} R^{\alpha\beta\sigma}_{\,\,\,\,\,\,\,\,\,\,\nu)}+ 4 f_{P} R^{\lambda}_{\;\;(\mu}R_{\nu)\lambda}\bigg).
    \end{split}
\end{equation}
   This equation matches the stress-energy tensor found in \cite{frq} for $f(R,Q)$ gravity by setting $P=0$. Note, however, that the latter equation still carries a fourth-order derivative term, namely, $Y_{\mu\nu}= -\left(f_{Q}+4f_{P}\right)\Box R_{\mu\nu}$, in contrast to what is claimed in \cite{barr}. On the other hand, all components of such a higher-derivative term vanish altogether in the co-frame (\ref{eq:Lor_tetr}) for the special class of G\"{o}del-type correspondent to $m^2=4\omega^2$, as we can explicitly see below
   \begin{equation}
       Y_{(0)(0)}=4\left(f_{Q}+4f_{P}\right)(4\omega^2-m^2)\omega^2,\;\;\; Y_{(1)(1)}=Y_{(2)(2)}=2\left(f_{Q}+4f_{P}\right)(4\omega^2-m^2)\omega^2,
   \end{equation}
thus, for this specific case, the field equations reduce to a couple of second-order partial differential equations. Another possibility would be to enforce the constraint $f_{Q}=-4f_{P}$. Before proceeding any further, it is worth stressing that we shall scrutinize G\"{o}del-type solutions belonging to the three classes -- hyperbolic, linear and trigonometric -- differently from \cite{barr}, where the authors just explored the hyperbolic class. 

\subsection{Vacuum solutions}

We here are interested in investigating solutions of the gravitational field equations in the absence of matter sources, $T_{\mu\nu}^{(m)}=0$, boosted in parts by the results found in \cite{frq} in the context of $f(R,Q)$ gravity. As a first step, let us proceed with the shift in $f$, namely, $f(R,Q)=f(R,Q)-2\Lambda$, in order to implement the cosmological constant for the sake of completeness. Formally, the insertion of a cosmological constant can be viewed as an effective contribution on the \textit{r.h.s} of the field equations (\ref{eqw}), more explicitly, such a contribution generates a stress-energy tensor given by $T^{(\Lambda)}_{AB}=-\Lambda\eta_{AB}$.

Taking this into account, the field equations (\ref{eqfi}) in the local Lorentz basis (\ref{eq:Lor_tetr}) read
\begin{eqnarray}
   \label{eq1}0&=& 4\omega^2 f_{R}-f+2\Lambda -16 f_{Q}\omega^2\left(3\omega^2 - m^2\right)+16f_{P}\omega^2 \left(3m^2 -11\omega^2\right);\\
   \nonumber \label{eq20}0&=&2f_{R}\left(2\omega^2 - m^2\right) +f-2\Lambda-4f_{Q}\left(m^4 -6\omega^2 m^2 +12\omega^4\right)-\\
    &-&8f_{P}\left(m^4 -9\omega^2 m^2 +22\omega^4\right);\\
   \label{eq3}2\Lambda&=&f,
\end{eqnarray}
in addition, we have the constraint (\ref{cons})
\begin{equation}
    f_{R}R+2f_{Q}Q+2f_{P}P-2f+4\Lambda=0.
    \label{eq4}
\end{equation}
Upon combining Eqs. (\ref{eq1}, \ref{eq20}, \ref{eq3}, \ref{eq4}), we are able to find the relation 
\begin{equation}
    f_{R}-4f_{Q}\left(3\omega^2 -m^2\right)+4f_{P}\left(3m^2 -11\omega^2\right)=0.
    \label{ews}
\end{equation}
This equation leads us to some important conclusions. First, by taking $f_{Q}=f_{P}=0$, we obtain that $f_{R}=0$; thereby one concludes that $f(R)$ theories do not admit G\"{o}del-type solutions in the absence of matter sources. As an immediate consequence of that fact, GR also does not. Second, by assuming $f_{P}=0$, in other words, $f=f(R,Q)$, we recover the results found in \cite{frq}.  Another important point to mention is that our results disagree with those found in \cite{barr} for the hyperbolic class of G\"{o}del-type metrics in the absence of matter sources. Conversely, as we have pointed out before, our results are in agreement with  
those found in \cite{frq,RT2} for  $f(R,Q)$ and $f(R)$ theories of gravity, respectively. One can get some general results by exploring further Eq.(\ref{ews}). For example, to accomplish completely causal G\"{o}del-type metrics (without the presence of CTCs), which corresponds to $m^2\geq 4\omega^2>0$, Eq. (\ref{ews}) must satisfy the following inequality amid $f_{R}, f_{Q}$ and $f_{P}$, i.e.,
\begin{equation}
    f_{R}\le -4\omega^2 \left(f_{Q}+f_{P}\right).
\end{equation}
So, one concludes that theories  fulfilling the previous inequality do not present CTCs. 
For the linear class ($m^2=0$), the relation amongst $f_{R}, f_{Q}$ and $f_{P}$ must satisfy
\begin{equation}
    f_{R}= 4\omega^2 \left(3f_{Q}+11f_{P}\right),
\end{equation}
with the critical radius ($r_{c}=1/\omega$) given by
\begin{equation}
    r_{c}=2\sqrt{\frac{3(f_{Q}+11f_{P})}{f_{R}}}.
\end{equation}
On the other hand, the trigonometric class ($m^2=-\mu^2$) is achieved when
\begin{equation}
    \frac{12 f _{Q}\omega^2 +44f_{P}\omega^2-f_{R}}{4\left(f_{Q}+3f_{P}\right)}<0.
\end{equation}

To shed more light on the vacuum solutions, Eq. (\ref{ews}), it should be demanded the explicit knowledge of the function $f$. In order to do so, let us treat some particular functional forms of $f$ now: 

\subsubsection{$f(R,Q,P)=R+\alpha R^2 +\beta Q + \gamma P$}
In this particular case, Eq.(\ref{ews}) reduces to
\begin{equation}
    4m^2 = \frac{4\omega^2 \left(\alpha+3\beta+11\gamma\right)-1}{\alpha +\beta+3\gamma},
\end{equation}
where $\alpha, \beta$ and $\gamma$ are constant parameters.
Of course, the case $\alpha=\beta=\gamma=0$, which corresponds to GR, is precluded, as expected. However, by introducing the Gauss-Bonnet scalar invariant, $\mathcal{G}=R^{2}-4Q+P$, which is a topological invariant (total derivative) in four dimensions,  this particular model $f(R,Q,P)=R+\alpha R^2 +\beta Q + \gamma P$ can be mapped into $f(R,Q)=R+\alpha^{\prime} R^2 +\beta^{\prime} Q$ by means of a redefinition of their parameters. Such theory has been already explored in \cite{Ac} and completely causal vacuum solutions have been found there, apart from other results. Another plausible possibility would be to consider $f(R,Q,P)=R+f(\mathcal{G})$, where $f(\mathcal{G})$ corresponds to an arbitrary function of Gauss-Bonnet scalar invariant \cite{MontelongoGarcia:2010ip}. In this situation, we have $f_{R}=1+2f_{\mathcal{G}}R$, $f_{Q}=-4f_{\mathcal{G}}$ and $f_{P}=f_{\mathcal{G}}$. Plugging these identities into Eq. (\ref{ews}), one finds an inconsistency that entails that this model does not admit G\"{o}del-type solutions. Of course, this result also holds for $f(R,Q,P)=F(R,\mathcal{G})$ generalized gravity models since $f_{R}=F_{R}+2F_{\mathcal{G}}R$, $f_{Q}=-4F_{\mathcal{G}}$ and $f_{P}=F_{\mathcal{G}}$ do not fulfill Eq.(\ref{ews}), except for $F_{R}=0$, which is not a realistic case because does not recover GR at the low-energy limit. These models have been recently used to explain the cosmic speed-up of the Universe \cite{OOB}.

\subsubsection{$f(R,Q,P)=R+g(a R^2+b Q+c P)$}

This model generalizes the previous case by considering a generic function of the higher order curvature invariants, $g(a R^2+b Q+c P)$. Actually, such a class of general theories has been explored in several contexts \cite{Carroll:2004de}. To have more explicit results and for the sake of convenience, we shall treat the particular class of theories defined by 
\begin{equation}
    g(a R^2+b Q+c P)=-\frac{\mu^{4n+2}}{\left(a R^2+b Q+c P\right)^n},
    \label{jk0}
\end{equation}
where $a, b$ and $c$ are dimensionless constant parameters, $n$ is a positive integer number and $\mu$ is a mass scale constant. In particular, the choice (\ref{jk0}) is useful to study modifications of the Einstein-Hilbert action in the infrared regime that means that such modifications manifest relevance at low curvatures. This scenario is important in cosmology, for instance, to investigate the late-time accelerating cosmic expansion \cite{Carroll:2004de}.  

All three classes of G\"{o}del-type metrics are possible for this specific class of theory. We remark that the first completely causal solution ($m^2 =4\omega^2$) must satisfy
\begin{equation}
    \omega^{4n+2}=-\frac{n\mu^{4n+2}}{ 12^{n}\cdot 3\left(3a+b+c\right)^n},
\end{equation}
with the additional conditions: $n$ being a positive odd number and $3a+b+c<0$ to assure the positivity of the vorticity. The linear class is also a solution if 
\begin{equation}
    \omega^{4n+2}=\frac{n\mu^{4n+2}}{4^{n}\left(a+3b+11c\right)^n}.
\end{equation}
Note that this solution supports either $a+3b+11c>0$ or $a+3b+11c<0$ when $n$ being a positive even number. In contrast to $n$ being a positive odd number, which only supports $a+3b+11c>0$.

Another interesting model consists of taking $b=-4c$ in Eq. (\ref{jk0}) so that the Gauss-Bonnet term $\mathcal{G}$ emerges. In this case, 
\begin{equation}
g(aR^2 +bQ+cP)=g\left((a-c)R^2+c\mathcal{G}\right)=-\frac{\mu^{4n+2}}{\left((a-c)R^2+c\mathcal{G}\right)^n},
\end{equation}
which leads to the following completely causal solution ($m^2=4\omega^2$)
\begin{eqnarray}
    \omega^{4n+2}=-\frac{n\mu^{4n+2}}{ 12^{n}\cdot 3^{n+1}\left(a-c\right)^n},
\end{eqnarray}
with the further conditions: $n$ being a positive odd number and $a-c<0$, similarly to the aforementioned general case. For the linear class, we achieve the following solution
\begin{equation}
    \omega^{4n+2}=\frac{n\mu^{4n+2}}{4^{n}\left(a-c\right)^n},
\end{equation}
where should be fulfilled the following requirements: if $n$ is odd then $a-c>0$ and if $n$ is even then either $a-c>0$ or $a-c<0$. Such conditions guarantee the positivity of the vorticity.

\subsection{Solutions in the presence of matter sources}

Here we introduce a new ingredient: the matter sources. Altogether, let us deem a perfect fluid and a massless scalar field as the matter content, in much the same way as \cite{frq}. The stress-energy tensor of a perfect fluid is defined by $T_{AB}^{(pf)}=(p+\rho)u_{A}u_{B}-p\eta_{AB}$, in the tetrad basis (\ref{eq:Lor_tetr}), where $p$ and $\rho$ are the pressure and density of the perfect fluid, respectively. The $4$-velocity of a particular point particle of this perfect fluid is characterized by $u^{A}=\delta^{A}_{0}$. Putting all this information together, we arrive at the only non-zero components of the stress-energy tensor are 
\begin{equation}
    T_{(0)(0)}^{(pf)}=\rho, \;\; T_{(1)(1)}^{(pf)}=T_{(2)(2)}^{(pf)}=T_{(3)(3)}^{(pf)}=p. 
\end{equation}
The second matter source is a massless scalar field ($\psi$) in which its stress-energy tensor, in the local Lorentz co-frame (\ref{eq:Lor_tetr}), $T_{AB}^{(sf)}=\nabla_{A}\psi\nabla_{A}\psi-\frac{1}{2}\eta_{AB}\eta^{CD}\nabla_{C}\psi\nabla_{D}\psi$, where $\nabla_{A}\psi=e^{\;\;\mu}_{A}\nabla_{\mu}\psi$ as usual. In addition, the scalar field must satisfy the Klein-Gordon equation, $\Box\psi=\eta^{AB}\left(\nabla_{A}\nabla_{B}\psi+\omega^{C}_{\;BA}\nabla_{C}\psi\right)=0$, where $\omega^{C}_{\;BA}$ are the non-holonomic coefficients. Using the symmetries of the G\"{o}del-type metrics, it is reasonable to restrict the scalar field to be $z$-dependent only. In this case, by solving the Klein-Gordon equation, we obtain $\psi=bz+b_{0}$, where $b$ and $b_{0}$ are constants. It is worth calling attention to the fact that the gradient of the scalar field, $\nabla_{A}\psi=[0,0,0,b]$, is aligned along the same direction, at each point of the space-time, as the angular velocity, $\omega^{A}=\frac{1}{2}\epsilon^{ABCD}\omega_{BCD}=[0,0,0,\omega]$. Thus, the non-zero components of the stress-energy  tensor of this scalar field are given by
\begin{equation}
    T_{(0)(0)}^{(sf)}=T_{(3)(3)}^{(sf)}=\frac{1}{2}b^2, \;\; T_{(1)(1)}^{(sf)}=T_{(2)(2)}^{(sf)}=-\frac{1}{2}b^2. 
\end{equation}
  Assembling both matter sources, we have $T_{AB}^{(m)}=T_{AB}^{(pf)}+T_{AB}^{(sf)}$ whose non-zero components in the non-local Lorentz co-frame (\ref{eq:Lor_tetr}) are given by
  \begin{equation}
      T_{(0)(0)}^{(m)}=\rho+\frac{1}{2}b^2,\;\; T_{(1)(1)}^{(m)}=T_{(2)(2)}^{(m)}=p-\frac{1}{2}b^2,\;\; T_{(3)(3)}^{(m)}=p+\frac{1}{2}b^2.
  \end{equation}

  The non-vanishing components of the modified Einstein equations (\ref{eqfi}), in the presence of the aforementioned matter content, look like
  \begin{eqnarray}
   \nonumber 0&=& 4\omega^2 f_{R}-f -16 f_{Q}\omega^2\left(3\omega^2 - m^2\right)+16f_{P}\omega^2 \left(3m^2 -11\omega^2\right)-\\
   \label{eq11}&-&2\kappa^2 \rho -\kappa^2 b^2;\\
   \nonumber \label{eq2}0&=&2f_{R}\left(2\omega^2 - m^2\right) +f-4f_{Q}\left(m^4 -6\omega^2 m^2 +12\omega^4\right)-\\
    \label{eq22}&-&8f_{P}\left(m^4 -9\omega^2 m^2 +22\omega^4\right)-2\kappa^2 p +\kappa^2 b^2;\\
   \label{eq33}0&=&f-2\kappa^2 p -\kappa^2 b^2.    
  \end{eqnarray}
Solving the previous set of algebraic equations in terms of the matter sources, we found that
\begin{eqnarray}
    \nonumber\kappa^2 b^2&=&(m^2 - 2\omega^2)f_{R}+2(m^4 -6\omega^2 m^2 +12\omega^2)f_{Q}+\\
    \label{eqm1}&+&4(m^4 +9\omega^2 m^2 -22\omega^4 )f_{P};\\
    \nonumber \kappa^2 p&=& \frac{1}{2}f -\frac{1}{2}(m^2 - 2\omega^2)f_{R}-(m^4 -6\omega^2 m^2 +12\omega^2)f_{Q}-\\
    \label{eqm2}&-&2(m^4 +9\omega^2 m^2 -22\omega^2)f_{P};\\
    \nonumber \kappa^2 \rho &=& -\frac{1}{2}f -\frac{1}{2}(m^2 - 6\omega^2)f_{R}-(m^4 -14\omega^2 m^2 +36\omega^2)f_{Q}-\\
    \label{eqm3}&-&2(m^4 -3\omega^2 m^2 +22\omega^2)f_{P}.
\end{eqnarray}   
Making use of the above equations one can obtain information on the causality properties of the G\"{o}del-type metrics. First, It realizes that by taking $f_{P}=0$, which corresponds to $f(R,Q,P)=f(R,Q)$, we recover the results found in \cite{frq} and, particularly, those ones found in \cite{Ac} for $f(R,Q)=\alpha R+\beta Q$. On the other hand, let us take the particular class of the G\"{o}del metric ($m^2=2\omega^2$), which presents CTCs, to illustrate the causality properties of the general case, i.e., when $f_{P}\neq 0$. In this scenario, the critical radius, whose explicit form is given by Eq. (\ref{eq:raiocr}), looks like
\begin{equation}
    r_{c}=\frac{2}{m}\sinh^{-1}{(1)}=4\sqrt{5}\sinh^{-1}{(1)}\sqrt{{\frac {f_{P}}{ \left( f_{R}+\sqrt {{f_{R}}^{2}-40\,\kappa^{2} \left( 2\,{b}^{2}+p+\rho \right) f_{P}} \right) }}},
    \label{er}
\end{equation}
where we have made use of Eqs. (\ref{eqm1}, \ref{eqm2}, \ref{eqm3}) in the second equality. Recalling that we must assume the following requirements: ${f_{R}}^{2}\geq 40\,\kappa^{2}\left( 2\,{b}^{2}+p+\rho \right) f_{P}$ and $f_{P}> 0$, so that the critical radius equation (\ref{er}) holds. Similarly, one can compute the critical radius for the linear class ($m^2=0$), using Eqs. (\ref{eqm1}, \ref{eqm2}, \ref{eqm3}) besides Eq. (\ref{erf}), thereby
\begin{equation}
    r_{c}=22\left[\frac{f_{P}}{\kappa^{2}(2b^2 +p+\rho)}\right]^{1/4},
\end{equation}
with either $f_{P}>0$ and $2b^2 +p+\rho<0$ or $f_{P}<0$ and $2b^2 +p+\rho>0$ being satisfied. 

\subsubsection{{\bf Completely causal solutions}}

As previously discussed, the metric parameters must satisfy the inequality $m^2\geq 4\omega^2$ in order to circumvent the arising of CTCs. We here analyze the conditions on the gravitation field equations in the presence of matter sources (a perfect fluid and scalar field) to engender completely causal solutions.

Our starting  point is rewriting the field equations (\ref{eq11}, \ref{eq22}) and (\ref{eq33}) in a more convenient form, i.e.,
\begin{eqnarray}
    \nonumber 2f_{R}&=& \frac{8\left(264\omega^6 +41m^4 \omega^{2}-200m^{2}\omega^{4}-m^6 \right)f_{P}}{m^2 (4\omega^2 -m^2)}+\frac{\kappa^2\left(m^4 -6\omega^2 m^2 +12\omega^4\right)\left(p+\rho\right) }{\omega^2 m^2 (4\omega^2 -m^2)}\\
    \label{ww} &+&\frac{\kappa^2 \left(6\omega^2 -m^2\right)b^2}{\omega^2 m^2} ;\\
    \label{www}4f_{Q}&=& \frac{16\left(2\omega^2 -m^2\right)\left(11\omega^2 - m^2\right)f_{P}}{m^2 (4\omega^2 - m^2)}+\frac{\kappa^2 \left(2\omega^2 -m^2\right)\left(\rho+p\right)}{\omega^2 m^2 \left(4\omega^2 -m^2\right)}+\frac{\kappa^2 b^2}{\omega^2 m^2};\\
    f&=&\kappa^2 \left(2p+b^2\right),
\end{eqnarray}
with $m^2 \neq 4\omega^2$, $m^2\neq 0$ and, of course, $\omega^2\neq 0$. Nonetheless, for the class $m^2=4\omega^2$, which corresponds to the first completely causal solution, we have
\begin{eqnarray}
    \kappa^2 (p+\rho)&=&-112\omega^4 f_{P};\\
    \label{jk}\kappa^2 b^2 &=&120\omega^4f_{P}+8\omega^4f_{Q} + 2 \omega^2 f_{R};\\
    \kappa^2 (b^2 +2p)&=&f.
\end{eqnarray}
It realizes from the above equations that the pure scalar field case cannot generate the class $m^2=4\omega^2$ whether $f_{P}\neq 0$. As $f_P =0$ one recovers the same results obtained in \cite{frq}.  On the other hand,  the pure perfect fluid case constraints the functional form of $f(R,Q,P)$, as we can see from Eq. (\ref{jk}) that reduces to 
\begin{equation}
f_R=-60\omega^2 f_P - 4\omega^2 f_Q.
\label{kj}
\end{equation}
 This suggests that the generic case permits new completely causal solutions within the range $m^2>4\omega^2$ and also a further arbitrariness on the functional form of $f(R,Q,R)$.

In order to investigate such CTC-free solutions, let us assume $m^2>4\omega^2$ in Eqs. (\ref{ww}, \ref{www}). Note yet that these solutions can be broken into two distinct cases:

\begin{itemize}
    \item $p+\rho>0$ and $b^2>0$. 

    This case is achieved if the conditions below are satisfied:
    
    $
       \begin{cases}
    f_{Q}>0, & \mbox{if} \,\, 4\omega^2<m^2<11\omega^2 \,\, \mbox{and}\,\, f_{P}>0\\
    &\mbox{or} \,\, m^2>11\omega^2\,\, \mbox{and}\,\, f_{P}<0, \\
             f_{R}<0, & \mbox{if}\,\, 6\omega^2\leq m^2 \lesssim 35.5887\omega^2 \,\, \mbox{and} \,\, f_{P}>0\\
             &\mbox{or} \,\, m^2\gtrsim 35.5887\omega^2\,\, \mbox{and}\,\, f_{P}<0.
       \end{cases}
$
\item $p+\rho<0$ and $b^2>0$.

In this scenario, we conclude from Eqs. (\ref{ww}, \ref{www}) that $f_{R}$ and $f_{Q}$ can assume both signs regardless of the sign of $f_{P}$.
\end{itemize}

To illustrate the aforementioned discussion, let us consider again the particular theory 
\begin{equation}
f(R,Q,P)=R -\frac{\mu^{4n+2}}{\left(a R^2+b Q+c P\right)^n}.
\end{equation}
 For convenience and without loss of generality, we shall concentrate our efforts on the simple case $n=1$ (the analysis is similar for generic $n$). It is straightforward from Eq.(\ref{kj}) that the parameters of the model must fulfill the condition  
 \begin{equation}
     \label{ker}\left(\frac{\omega}{\mu}\right)^{6}=-\frac{1}{4}\frac{3a+15c+b}{(9a+b+3c)^2}\Longrightarrow 3(a+5c)+b<0,
 \end{equation}
  in order to admit the first completely causal solution ($m^2=4\omega^2$) supported by a pure perfect fluid.

\section{Summary and conclusions}

In this work, we have investigated the conditions to engender completely causal G\"{o}del-type solutions within $f(R,Q,P)$ theories of gravity for a well-motivated matter content composed of a perfect fluid and a massless scalar field. As a first step, we have demonstrated that the field equations for $f(R,Q,P)$ do not reduce to a set of second-order partial differential equations for G\"{o}del-type metrics, in disagreement with \cite{barr}. Actually, this just happens for the special class $m^2=4\omega^2$, which corresponds to an important and non-trivial result since it avoids the potential arising of ghosts and also the emergence of CTCs. Despite these models presenting this particular CTC-free solution, we were successful to generate other causal solutions inside the range $m^2>4\omega^2$. Apart from that, it is noteworthy that the three classes of G\"{o}del-type metrics have been obtained, and then generalizing the results found in \cite{barr}, which takes into account only the hyperbolic class.  

We have found many solutions without any parallel with GR. For example, we have shown that $f(R,Q,P)$ gravity with a non-trivial cosmological constant  admits vacuum G\"{o}del-type solutions (all three classes), in particular, completely causal ones.  In this context, we have treated two particular models, namely: $f(R,Q,P)=R+\alpha R^2 +\beta Q +\gamma P$ and $f(R,Q,P)=R -\frac{\mu^{4n+2}}{\left(a R^2+b Q+c P\right)^n}$. In the former case, this model is equivalent to $f(R,Q)=R+\alpha^{\prime} R^2 +\beta^{\prime} Q$, where the G\"{o}del-type metrics have been studied in \cite{Ac, barr}. In the latter case, however, it was provided the conditions on the constant parameters to accomplish the first CTC-free solution ($m^2=4\omega^2$).     

By including matter sources, a perfect fluid and/or a scalar field, we explicitly computed the critical radius $r_c$ for the G\"{o}del class ($m^2=2\omega^2$) and also for the linear class ($m^2=0$). We have also obtained the necessary conditions to achieve completely causal solutions in the presence of a perfect fluid and/or a scalar field. As a specific example, we have deemed the effective ghost-free model 
$f(R,Q,P)=R -\dfrac{\mu^{6}}{\left(a R^2+b Q+c P\right)}$. In particular, this theory presents a completely causal G\"{o}del-type solution if Eq.(\ref{ker}) holds. We close the paper with the statement that other forms of the function $f(R,Q,P)$ displaying similar behaviour also can exist.

\textbf{Acknowledgments.} This work was partially supported by Conselho
Nacional de Desenvolvimento Científico e Tecnológico (CNPq). PJP
would like to thank the Brazilian agency CNPq for financial
support (PQ–2 grant, process No. 307628/2022-1). The work
by A. Yu. P. has been supported by the CNPq project No. 301562/2019-9.

\end{document}